\documentclass{aastex6}

\usepackage{graphicx}
\usepackage{amsmath}

\begin{document}

\title{EBL Inhomogeneity and Hard-Spectrum Gamma-Ray Sources}
\author{Hassan Abdalla \altaffilmark{1,2} and Markus B\"{o}ttcher\altaffilmark{1}}

\altaffiltext{1}{Centre for Space Research, North-West University, Potchefstroom 2520, South Africa}
\altaffiltext{2}{Department of Astronomy and Meteorology, Omdurman Islamic University, Omdurman 382, Sudan}

\begin{abstract}
The unexpectedly hard very-high-energy (VHE; $E > 100$~GeV) $\gamma$-ray spectra of a few distant 
blazars have been interpreted as evidence for a reduction of the $\gamma\gamma$ opacity of the 
Universe due to the interaction of VHE $\gamma$-rays with the extragalactic background light (EBL)
compared to the expectation from our current knowledge of the density and cosmological evolution of
the EBL. One of the suggested solutions to this problem consisted of the inhomogeneity of the EBL.
In this paper, we study the effects of such inhomogeneities on the energy density of the EBL 
(which then also becomes anisotropic) and the resulting $\gamma\gamma$ opacity. Specifically, 
we investigate the effects of cosmic voids along the line of sight to a distant blazar. We find that
the effect of such voids on the $\gamma\gamma$ opacity, for any realistic void size, is only of the 
order of $\lesssim 1$~\% and much smaller than expected from a simple linear scaling of the $\gamma\gamma$
opacity with the line-of-sight galaxy under-density due to a cosmic void. 
\end{abstract}


\section{Introduction} \label{sec:intro}
Gamma rays from astronomical objects at cosmological distances with energies greater than the threshold 
energy for electron-positron pair production can be annihilated due to $\gamma\gamma$ absorption by 
low-energy extragalactic photons. The importance of this process for high-energy astrophysics was first 
pointed out by \cite{Nishikov62}. In particular, very-high-energy (VHE; $E > 100$~GeV) $\gamma$-ray emission 
from blazars is subject to $\gamma\gamma$ absorption by the extragalactic background light
(EBL), resulting in a high-energy cut-off in the $\gamma$-ray spectra of blazars
\citep[e.g.,][]{Stecker92}. The probability of absorption depends 
on the photon energy and the distance (redshift) of the source. Studies of intergalactic $\gamma\gamma$ absorption 
signatures have attracted further interest in astrophysics and cosmology due to their potential use to probe
the cluster environments of blazars \citep{SB15} and to estimate cosmological parameters \citep{BW15}. 
However, bright foreground emissions prevent an accurate direct measurement of the EBL \citep{HD01}. 
Studies of the EBL therefore focus on the predicted $\gamma\gamma$ absorption imprints and employ
a variety of theoretical and empirical methods 
\citep[e.g.,][]{Stecker69,Stecker92,Aharonian06,Franceschini08,Razzaque09,Finke10,Dominguez11a,Gilmore12}. 
All the cited works agree that the universe should be opaque (i.e., $\tau_{\gamma\gamma} \gtrsim 1$)
to VHE $\gamma$-rays from extragalactic sources at high redshift ($z \gtrsim 1$). 

Observations of distant ($z \gtrsim 0.5$) $\gamma$-ray blazars have been interpreted by some authors 
\citep[e.g.,][]{Albert08,Archambault14} as evidence that the universe may be more transparent to very high 
energy $\gamma$-rays than expected based on all existing EBL models. Furthermore, several studies have
found that, after correction for EBL absorption, the VHE $\gamma$-ray spectra of several blazars appear
to be unexpectedly hard (photon indices $\Gamma_{\rm ph} \lesssim 1.5$) and/or exhibit marginal hints of 
spectral upturns towards the highest energies \citep[e.g.,][]{Finke10,Furniss13}. These unexpected VHE 
signatures in the spectra of distant blazars --- although present only at marginal significance --- are 
currently the subject of intensive research. 
Possible solutions include the  
hypothesis that the EBL density is generally lower than expected from current models \citep{Furniss13}; 
the existence of exotic axion like particles (ALPs) into which VHE $\gamma$-rays can oscillate in the 
presence of a magnetic field, thus enabling VHE-photons to avoid $\gamma\gamma$ absorption \citep{Dominguez11b}; 
an additional VHE $\gamma$-ray emission component due to interactions along the line of sight of 
extragalactic ultra-high-energy cosmic rays (UHECRs) originating from the blazar \citep[e.g.,][]{EK10}; 
and EBL inhomogeneities. The idea of EBL inhomogeneities was considered by \cite{Furniss15}, who 
found tentative hints for correlations between hard VHE $\gamma$-ray sources and under-dense regions 
along the line of sight. They suggested a direct, linear scaling of the EBL $\gamma\gamma$ opacity with
the line-of-sight galaxy number density. The effect of EBL inhomogeneities 
on the $\gamma\gamma$ opacity was also investigated by \cite{KF16}. However, in that work, EBL inhomogeneities
were considered only as a modulation of the redshift dependence of the cosmic star formation rate,
without a detailed consideration of the geometrical effects of large-scale structure of the Universe.
Both \cite{Furniss15} and \cite{KF16} concluded that the possible reduction of the EBL $\gamma\gamma$ 
opacity due to inhomogeneities is likely negligible.  \\

In this paper, we investigate the effect of cosmological inhomogeneities on the energy density of
the EBL and the resulting $\gamma\gamma$ opacity  with a detailed calculation of the inhomogeneous
and anisotropic EBL in a realistic geometrical model setup. Specifically, we will consider the effect of 
cosmic voids
along the line of sight to a distant blazar and investigate the resulting inhomogeneous
and anisotropic EBL radiation field. In Section \ref{Model}, we describe the model setup and the
method used to evaluate the EBL characteristics and the resulting $\gamma\gamma$ opacity. The results 
are presented in Section \ref{Results}, where we also compare our results to a
simple linear scaling of the EBL $\gamma\gamma$ opacity with the line-of-sight galaxy count density
for the specific example of PKS 1424+240.
We summarize and discuss our results in Section \ref{Summary}.

\section{EBL in the Presence of a Cosmic Void}
\label{Model}

Our calculations of the inhomogeneous EBL are based on a modified version of the formalism presented in
\cite{Razzaque09}, considering only the direct starlight. The effect of re-processing of starlight by
dust has been included in \cite{Finke10} and leads to an additional EBL component in the mid- to far
infrared, which is neglected here. Since dust re-processing is a local effect, it will be affected by
cosmic inhomogeneities in the same way as the direct starlight contribution considered here.

\begin{figure}[ht]
\begin{center}
\includegraphics[scale=0.4]{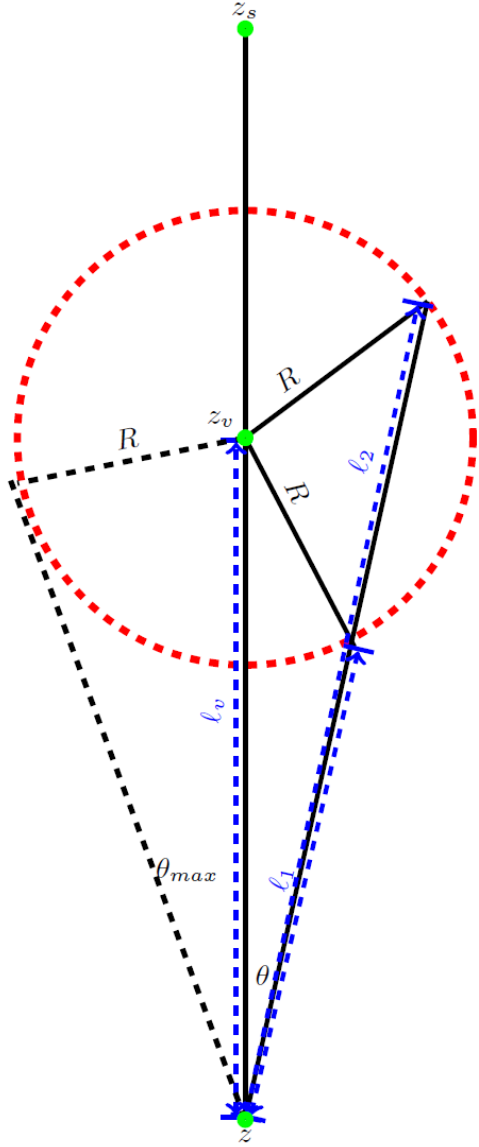} 
    \caption{ Illustration of an underdense region between the observer at redshift 
    $z$  and source at redshift $z_{s}$. We assume that the underdense region has a radius $R$ and 
    the redshift at the center of the underdense region is $z_{v}$.
    \label{fig:geometry}}
\end{center}
\end{figure}

For the purpose of a generic study of the effects of cosmic voids along the line of sight to a blazar,
we start out by considering a single spherical cosmic void
located with its center at redshift $z_v$ 
and radius $R$ between the observer and a $\gamma$-ray source at redshift $z_{s}$. The geometry is 
illustrated in Figure \ref{fig:geometry}.
We calculate the angle- and photon-energy-dependent EBL energy density at each point between the observer 
and the source by using co-moving cordinates, converting redshifts $z$ to distances $l (z)$. 
The cosmic void is represented by setting the star formation rate to 0 within the volume of the void.  

For the evaluation of the differential EBL photon number density spectrum at a given redshift $z$, we modify
the expression from \cite{Razzaque09}, based on the direct contribution from stars throughout cosmic history: 

\begin{equation}\label{ED}
\begin{split}
\frac{dN({\epsilon, z})}{d\Omega d\epsilon dV} = {\int_{\tilde z = z} ^{\infty}} d\tilde{z}  
\left|\frac{dt}{d \tilde{z}}\right|~ \Psi(\tilde{z}) f_{\rm void}(\Omega, \tilde{z}) ~ 
\int_{M_{min}} ^{M_{max}} dM \left( \frac{dN}{dM} \right) \\
~\times ~  \int_{max\{0,z_{d}(M,z^{'})\}} ^{\tilde{z}} dz^{'}\left|\frac{dt}{dz^{'}}\right| 
f_{\rm esc} (\epsilon^{'}) \frac{dN(\epsilon^{'}, M)}{ d\epsilon^{'} dt} (1+z^{'}).
\end{split}
\end{equation}
where $\Omega$ represents the solid angle with respect to the photon propagation direction, and 
$f_{\rm void}(\Omega, \tilde z)$ is the step function set to zero within the void, as specified below. 
$\Psi(\tilde{z})$
is the cosmic star formation rate, $dN/dM$ is the stellar mass function, $dN(\epsilon', M) / (d\epsilon' \, dt)$
is the stellar emissivity function, and $f_{\rm esc} (\epsilon')$ is the photon escape probability, 
for which we use the parameterizations of \cite{Razzaque09}. $dt/d \tilde{z}$ is evaluated using a concordance
cosmology with $\Omega_m = 0.3$, $\Omega_{\Lambda} = 0.7$ and $h = 0.7$, and we define $\tilde{\ell}$ as the
co-ordinate distance between $z$ and $\tilde{z}$.

From Figure \ref{fig:geometry} we can find the distances $l_{1}$ and $l_{2}$ where the gamma-ray propagation
direction $\Omega$ crosses the boundaries of the void, as 

\begin{equation}
\ell_{1,2} = \ell_v ~ \mu ~ { \mp } ~ \sqrt{  R^2  - \ell_v^2 \sin^{2} \theta }, 
\end{equation}
where $\mu = \cos\theta$ is the cosine of the angle betwen the line of sight and the gamma-ray propagation
direction, $\Omega = (\theta, \phi)$. The maximum angle $\theta_{\rm max}$ at which the gamma-ray travel
direction still crosses the boundary at one point, is given by:

\begin{equation}\label{max}
\sin\theta_{\rm max} =  \frac{R}{\ell_{v}}.
\end{equation}
and the corresponding distance to the tangential point, $\ell _{\rm max}$, is given by

\begin{equation}
\ell_{\rm max} = \ell_{v} \, \cos\theta_{\rm max}.
\end{equation}

The void condition can now be written as 

\[ f_{\rm void}^{\rm outside} ~ = ~ 
 \bigg\{
  \begin{tabular}{cc}
   0 &~ ~ if  ~    $\ell_{1} ~  < ~  \tilde{\ell} ~ <  ~ \ell_{2}$ \\
   1 & ~ ~ else 
  \end{tabular}
\]
for points $z$ along the line of sight that are located outside the void, and 

\[ f_{\rm void}^{\rm inside} ~ = ~ 
 \bigg\{
  \begin{tabular}{cc}
   0 &~ ~ if  ~    $ \tilde{\ell} ~ <  ~ \ell_{2}$ \\
   1 & ~ ~ else 
  \end{tabular}
\]
for points $z$ along the line of sight located inside the void. Note that, although the star formation 
rate has been set to zero inside the void, the EBL is not zero there because of the contribution from 
the rest of the Universe outside the void. 

To calculate the EBL density in comoving cordinates, the photon energy and volume can be transformed as 
$\epsilon_{1} = \epsilon(1+z_{1})$ and $V_{1} = V/(1+z_1)^3$ respectively. Using equation (\ref{ED}), the 
EBL energy density can then be written as \citep{Razzaque09}:

\begin{equation}
\epsilon_{1} \mu_{\epsilon_{1}}(\epsilon_1, z_1, \Omega) =  (1+z_{1})^{4} ~ \epsilon^{2} ~ 
\frac{dN(\epsilon,z = z_{1})}{d\Omega~d\epsilon ~ dV}.
\label{epsilon_mu}
\end{equation}
With this expression for the EBL energy density, we can calculate the optical depth due to $\gamma\gamma$ 
absorption for a $\gamma$-ray photon from a source at redshift $z_{s}$ with observed energy $E$ as \citep{GS67}:

\begin{equation}\label{opa}
\begin{split}
\tau_{\gamma\gamma} (E,z_s)= c ~~ {\int_{0} ^{z_{s}}} d{z_{1}}  
\left|\frac{dt}{d {z_{1}}}\right| \oint d\Omega  \int_{0} ^{\infty} d\epsilon_{1} ~  
\frac{\mu_{\epsilon_{1}}(\epsilon_1, z_1, \Omega)}{\epsilon_{1}} ~ (1-\mu) \sigma_{\gamma \gamma}(s).
\end{split}
\end{equation}
The $\gamma~\gamma$ pair-production cross section $\sigma_{\gamma \gamma}(s)$ can be written as:

\begin{equation}
\sigma_{\gamma \gamma}(s) = \frac{1}{2} \pi r_{e}^2 (1-\beta_{cm}^2) \left[ (3 - \beta_{cm}^4) 
\ln \left( \frac{1+\beta_{cm}}{1-\beta_{cm}} \right) - 2 \beta_{cm}(2 - \beta_{cm}^2)\right] H 
\left( \frac{( 1 + z_1 ) E \epsilon_{1} (1 - \mu)}{2 (m_e c^2)^2} - 1 \right)
\label{sigmagg}
\end{equation}
where $r_{e}$ is the classical electron radius and $\beta_{cm} = (1-\frac{1}{s})^{1/2}$ is the electron-positron 
velocity in the center-of-momentum (c.m.) frame of the $\gamma\gamma$ interaction, $s = \frac{s_{0}}{2} (1 - \cos \theta)$ 
is the c.m. frame electron/positron energy squared, $s_{0} = \frac{\epsilon E}{m_e^2 c^4} $ and H is the Heaviside 
function, $H(x) = 1$ if $x \ge 0$ and $H(x) = 0$ otherwise, representing the threshold condition that pair production 
can only occur if ${\frac{( 1 + z_1 ) E \epsilon_{1} (1 - \mu)}{2 (m_e c^2)^2}} > 1 $.

In the case of a homogeneous and isotropic EBL (with which we will compare our results for the inhomogeneous
EBL case), equation (\ref{opa}) can be simplified using the dimensionless function $\bar{\varphi}$ defined by 
\cite{GS67}:
$$\bar{\varphi}[s_{0}(\epsilon)] = \int_{1} ^{s_{0}(\epsilon)} s \bar{\sigma}(s) ds, $$
where $\bar{\sigma}(s) = \frac{2 \sigma(s)}{\pi r_{e}^{2}}$ and $s_{0}(\epsilon) = E (1+z) \epsilon / m_{e} ^2 c^{4}$,
so that equation (\ref{opa}) reduces to equation (17) in \cite{Razzaque09}:
\begin{equation}\label{opah}
\begin{split}
\tau_{\gamma\gamma}^{\rm hom} (E,z)= ~  c ~ \pi r_e ^2 \left(\frac{m_{e}^2 c^{4}}{E} \right)^{2} 
{\int_{0} ^{z_{s}}} \frac{d{z_{1}}}{(1+z_{1})^{2}}   \left|\frac{dt}{d {z_{1}}}\right| ~ \times~  
\int_{{\frac{m_{e}^2 c^4}{E (1+z1)}}} ^{\infty} d\epsilon_{1} ~ 
\frac{\mu_{\epsilon_{1}}}{\epsilon_{1}^{3}} ~ \bar{\varphi}[s_{0}(\epsilon)].
\end{split}
\end{equation}

Knowing the optical depth $\tau_{\gamma\gamma}$, we can calculate the attenuation of the intrinsic 
photon flux $F_{\nu}^{int}$ as:

\begin{equation}\label{exp} 
F_{\nu}^{obs} = F_{\nu}^{int} e^{-\tau_{\gamma\gamma} (E,z)},
\end{equation}
where $F_{\nu}^{obs}$ is the observed spectrum.

\begin{figure}[ht]
\plottwo{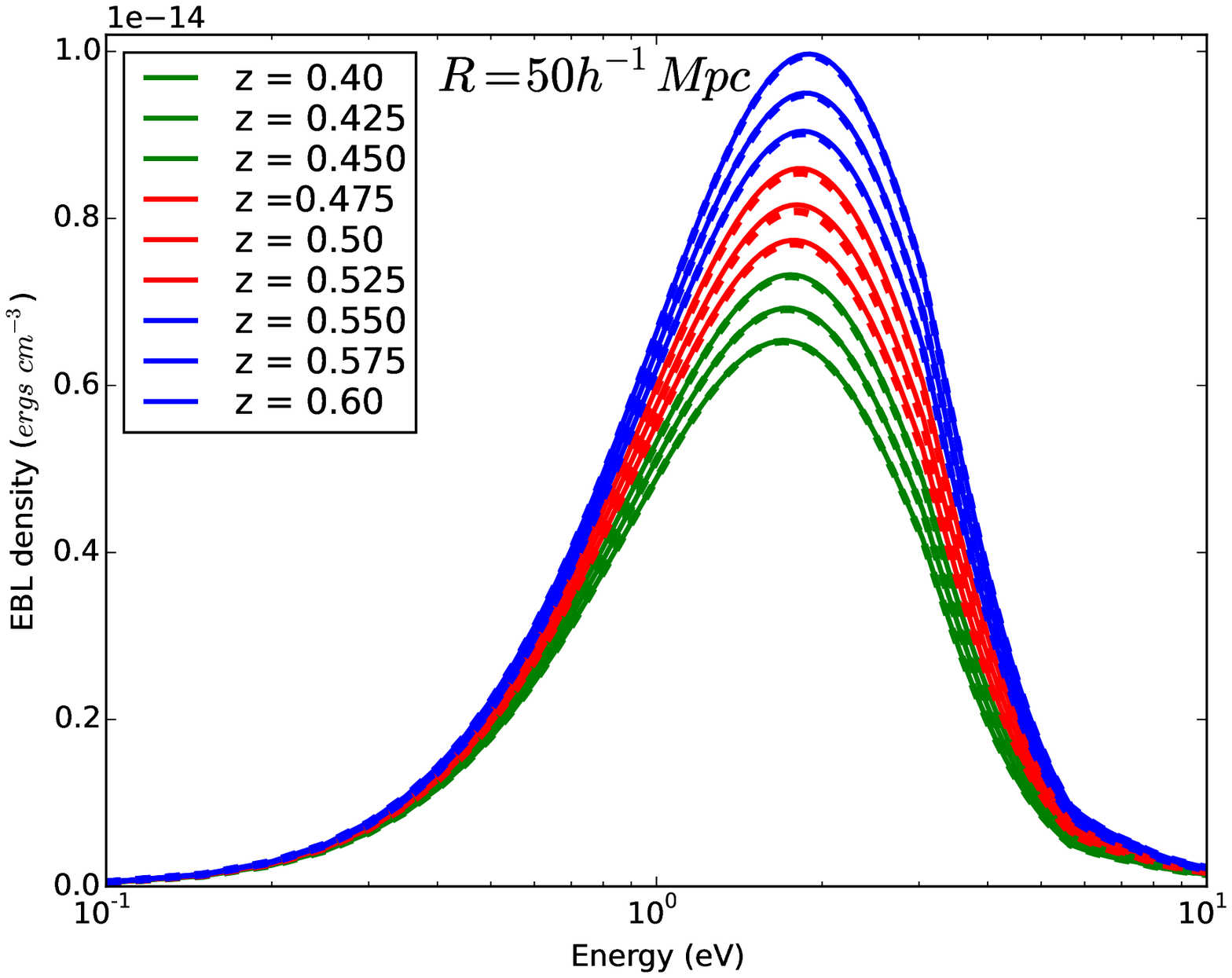}{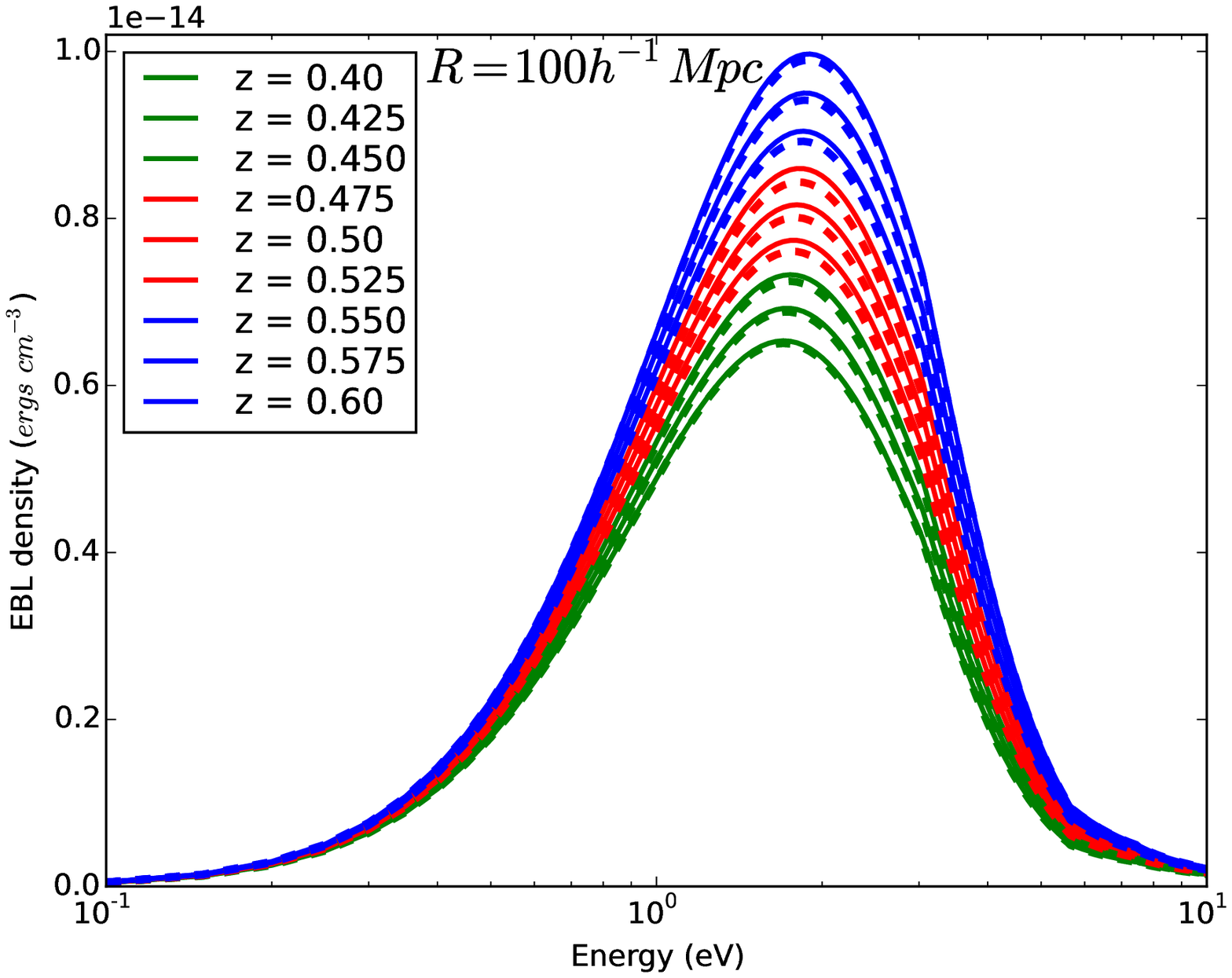}
\plottwo{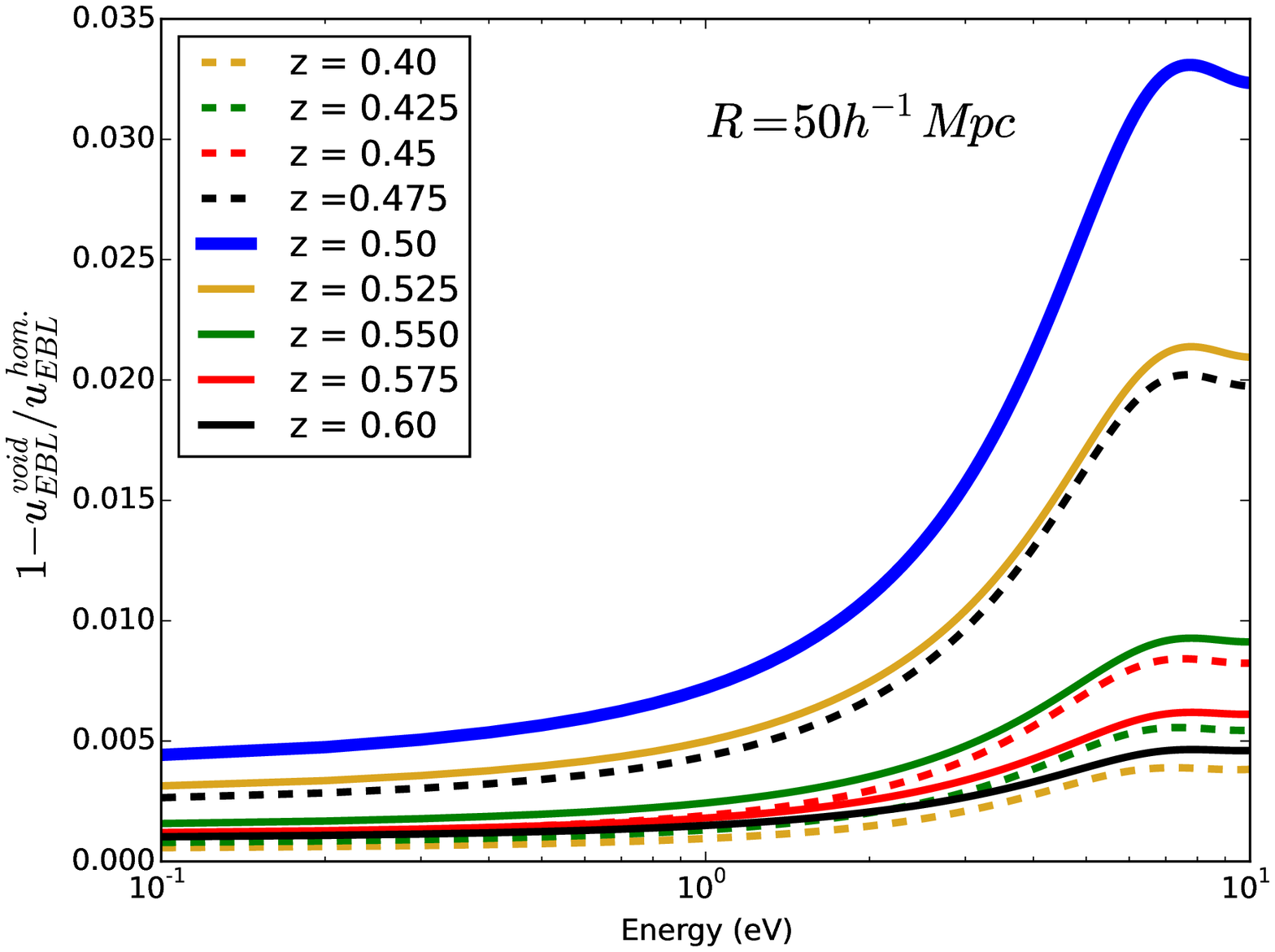}{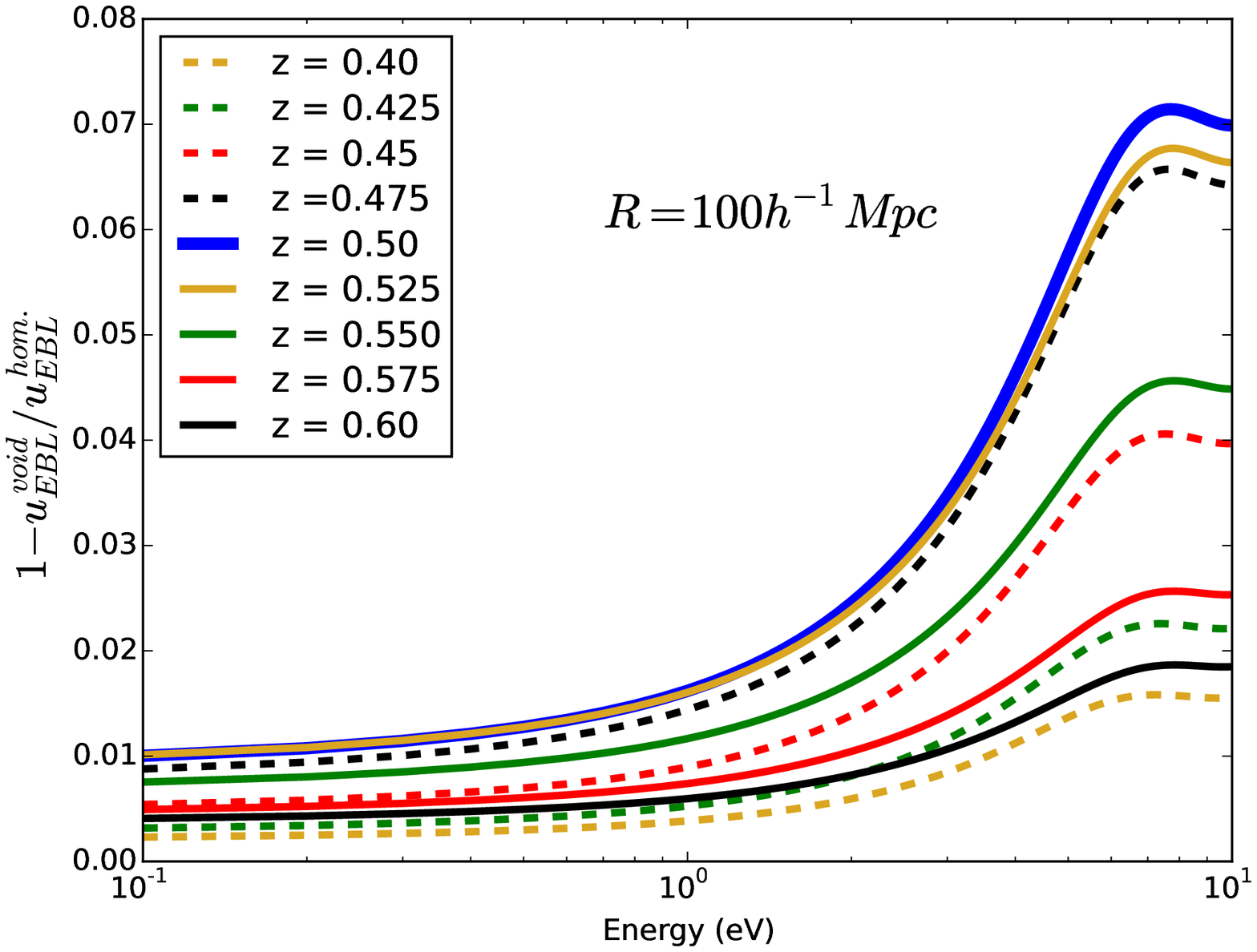}
\caption{Top panels: Angle-averaged EBL photon energy density spectra for a homogeneous EBL (solid lines) 
and in the presence of a spherical cosmic void (dahed lines) with its center at redshift $z_v = 0.5$ 
(comoving distance $1.724$~Gpc) and with radius $R = 50 \, h^{-1}$~Mpc (left) and $R = 100 \, h^{-1}$ (right). 
Green curves indicate locations in front of the void, red within the void, and blue behind the void. 
Bottom panels: Relative deficit of the EBL energy density due to the void. 
\label{fig:EBLdensities1}}
\end{figure}

\section{Results}
\label{Results}

\subsection{General Parameter Study: Single Void}
\label{parameterstudy}

We first investigate the effect of a single
cosmic void along the line of sight to a distant $\gamma$-ray source
on the resulting angle-averaged EBL energy density. Figure \ref{fig:EBLdensities1} (top panels) compares
the EBL energy density spectrum (keeping in mind that only the direct starlight contribution is accounted 
for) in the case of a void (dashed lines), compared to the homogeneous case (solid lines) for a spherical 
void of radius $R = 50 \, h^{-1}$~Mpc (left panels) and $R = 100 \, h^{-1}$~Mpc (right panels), at different
points (redshifts, as indicated by the labels) along the line of sight. The center of the void is assumed 
to be located at a redshift of $z_v = 0.5$, considering a source located at redshift $z_s \ge 0.6$. The
bottom panels of Figure \ref{fig:EBLdensities1} show the fractional difference between the homogeneous and
the inhomogeneous case as a function photon energy for various redshifts along the line of sight, for the
same two cases. As expected,
the effect of the void is largest right in the center of the void, but even there, 
it does not exceed a few \% (maximum fractional deficit $\sim 7$~\% in the $R = 100 \, h^{-1}$~Mpc case). 
The effect generally increases with photon energy. This is because high-energy photons are produced primarily 
by high-mass stars and, thus, trace the most recent star-formation history, which, for points within the 
void, is zero up to the time corresponding to the light travel time to the boundary of the void. As a 
function of position along the line of sight, the void-induced EBL deficit decreases approximately 
symmetrically for points in front of and behind the center of the void, with the slight asymmetry 
being due to the $(1 + z_1)^4$ factor in Equation \ref{epsilon_mu}.

\begin{figure}[ht]
\plottwo{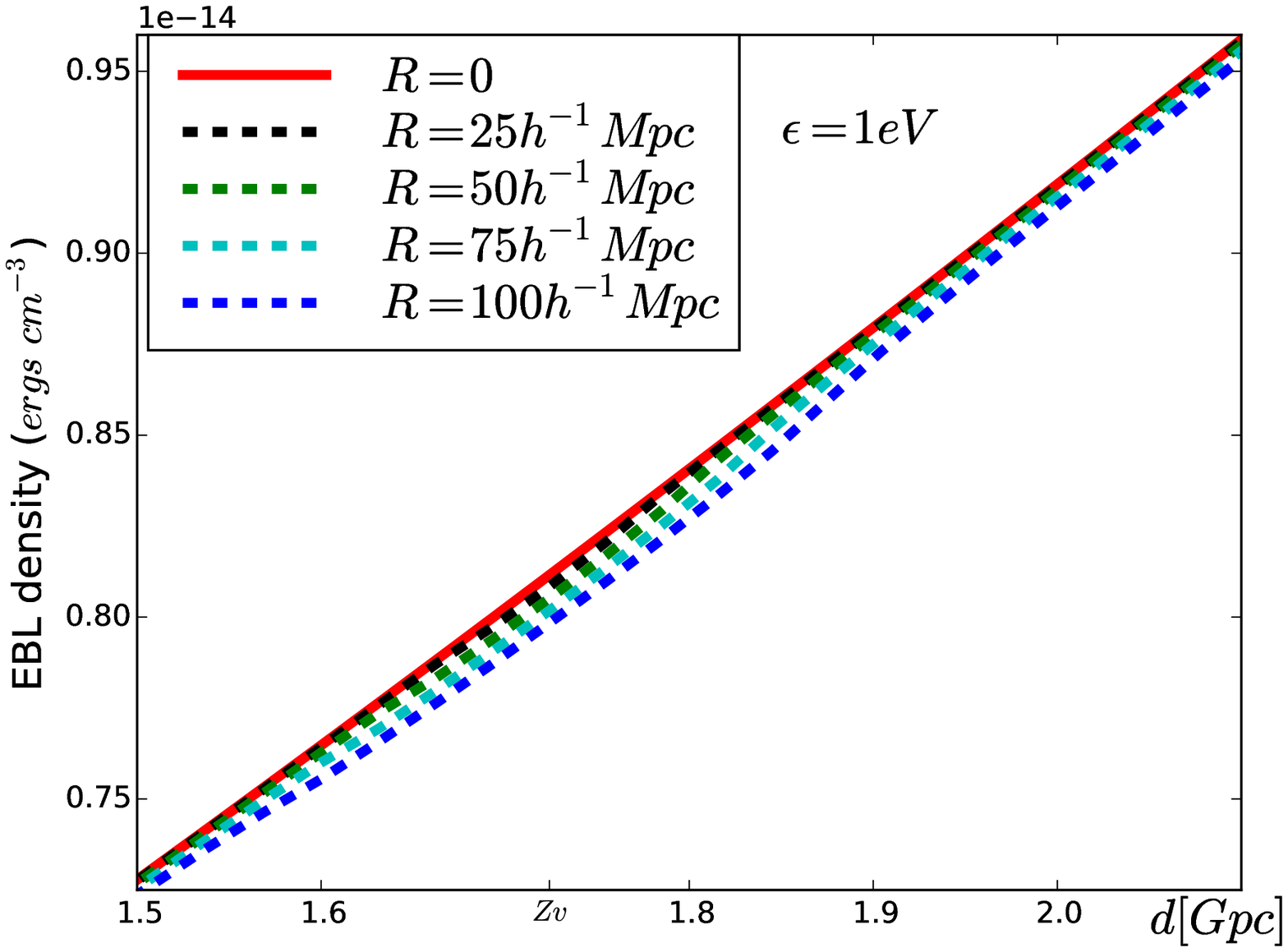}{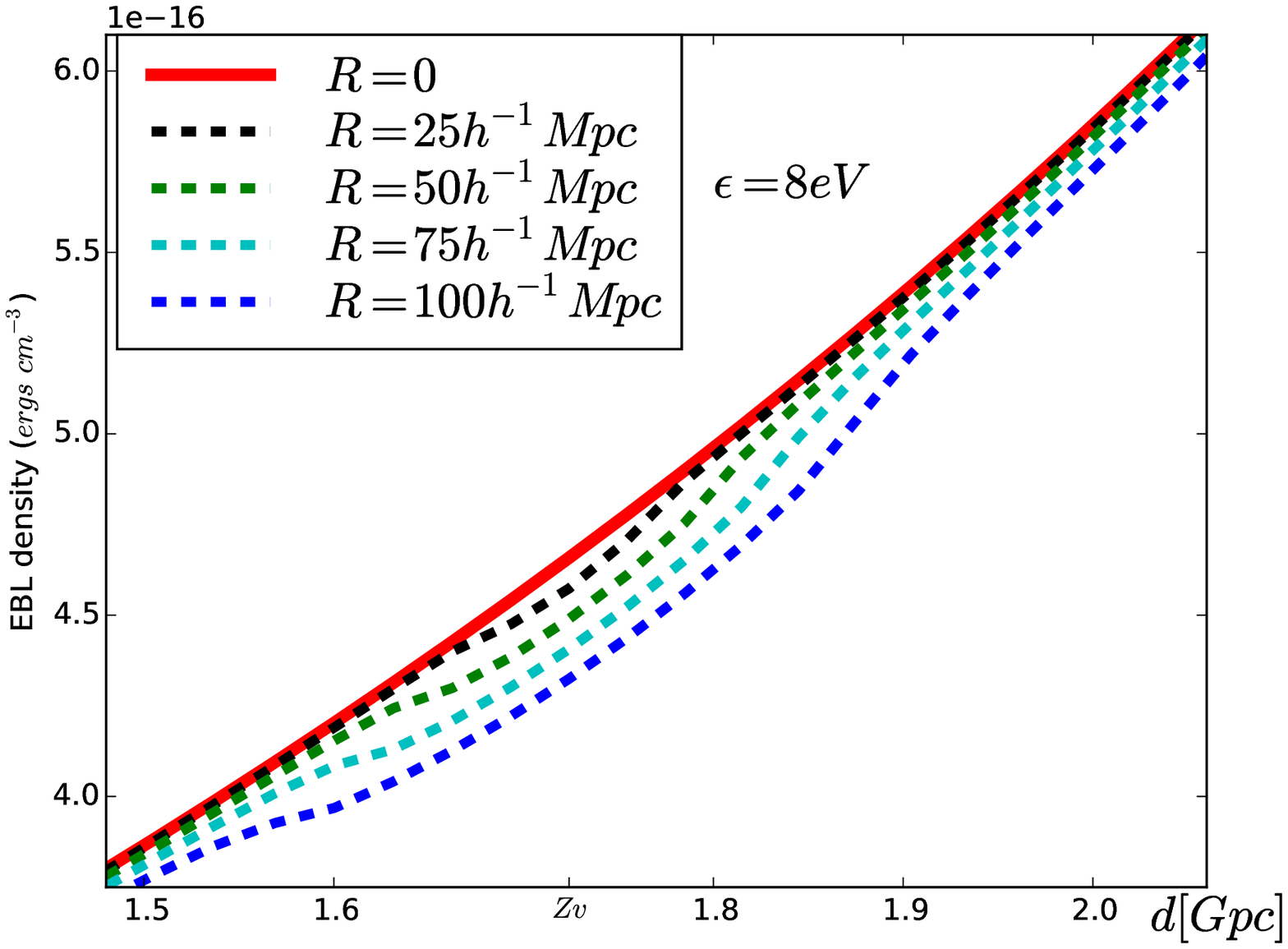}
\plottwo{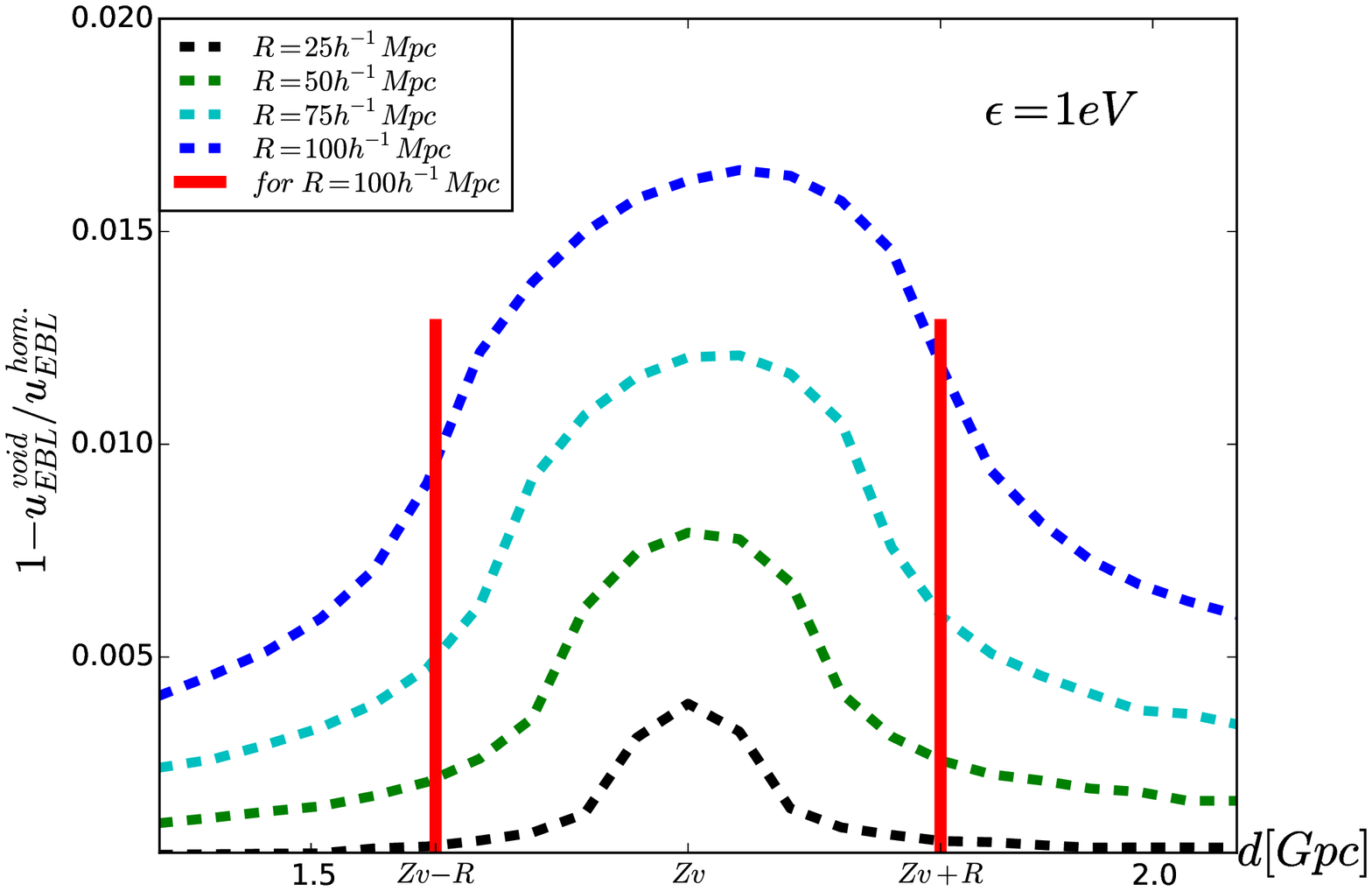}{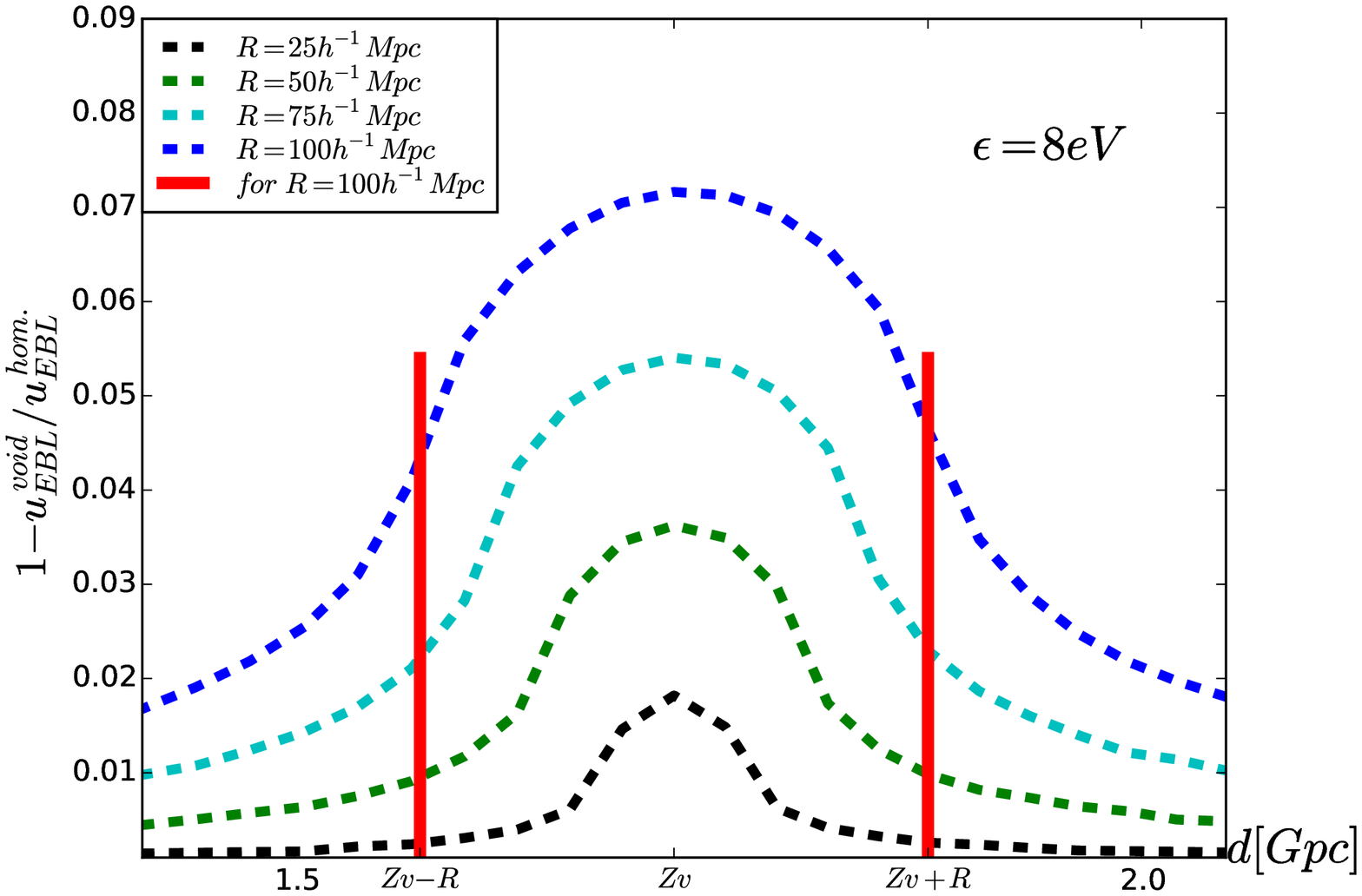}
\caption{Top panels: Differential EBL photon energy density as a function of distance along the line of
sight for various sizes (as indicated by the labels) of a void located at $z_v = 0.5$, at two representative
photon energies of $\epsilon = 1$~eV (left) and $\epsilon = 8$~eV (right). The red solid lines ($R = 0$) 
represent the homogeneous case. The general increase of the EBL energy density with redshift is due to the 
$(1 + z_1)^4$ factor in Equation \ref{epsilon_mu} and the increasing star formation rate with redshift. 
Bottom panels: Relativie EBL energy density deficit due to the presence of the void for the same cases
as in the top panels. The red vertical lines indicate the boundaries of the void for the 
$R = 100 \, h^{-1}$~Mpc case.
\label{fig:EBLdensities2} }
\end{figure}

Figure \ref{fig:EBLdensities2} illustrates the effect of the void on the differential EBL energy density
as a function of distance along the line of sight to the $\gamma$-ray source, for two representative EBL
photon energies, in the near-IR ($\epsilon = 1$~eV) and near-UV ($\epsilon = 8$~eV). 
The top panels show the absolute values of the energy densities, while the bottom panels show the 
fractional difference between the homogeneous and the inhomogeneous cases. The figure illustrates
that the maximum effect (at the center of the void) is approximately proportional to the size of the void,
but does not exceed $\sim 7$~\% in the case of the $R = 100 \, h^{-1}$~Mpc void.

\begin{figure}[ht]
\plottwo{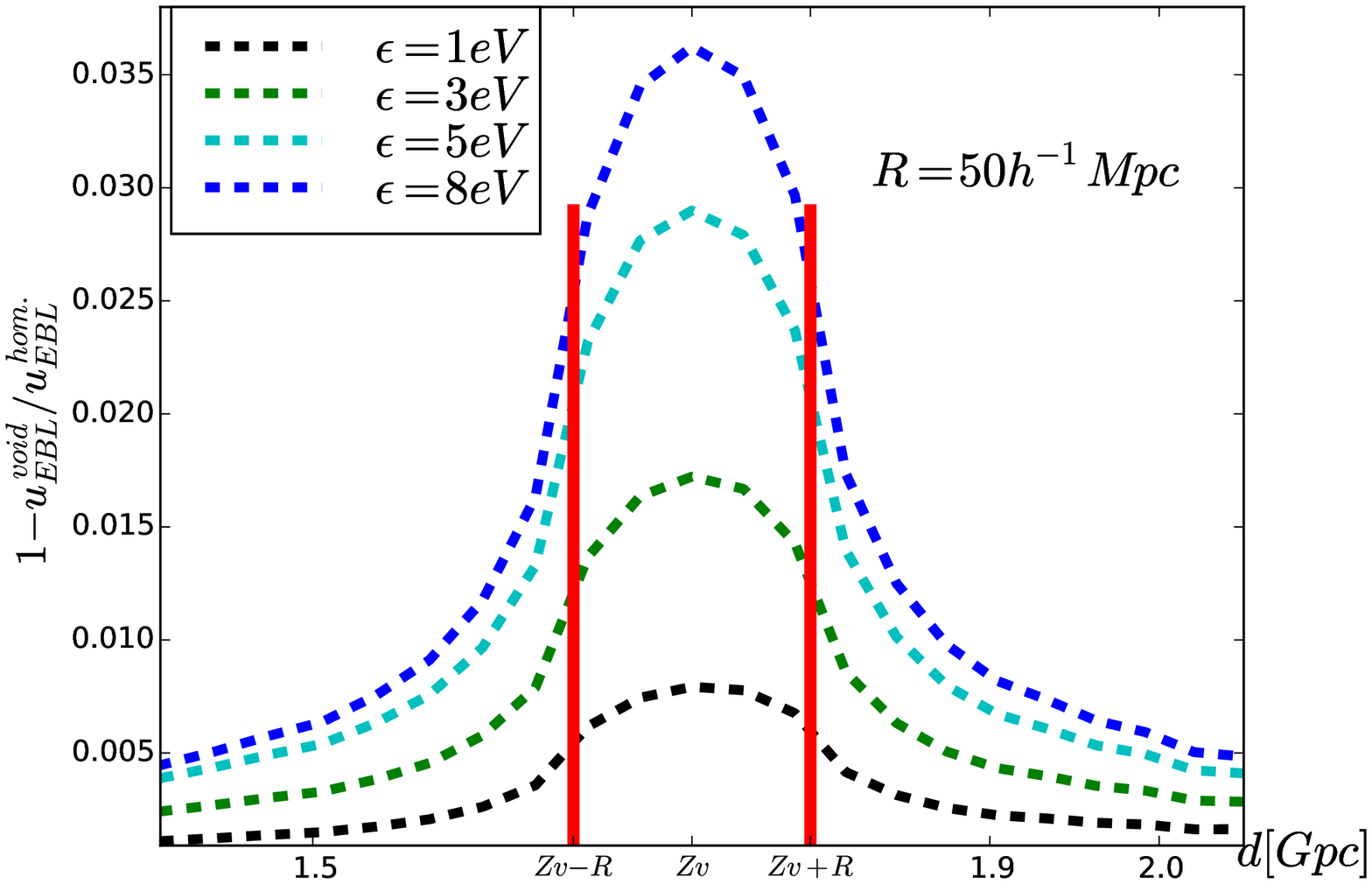}{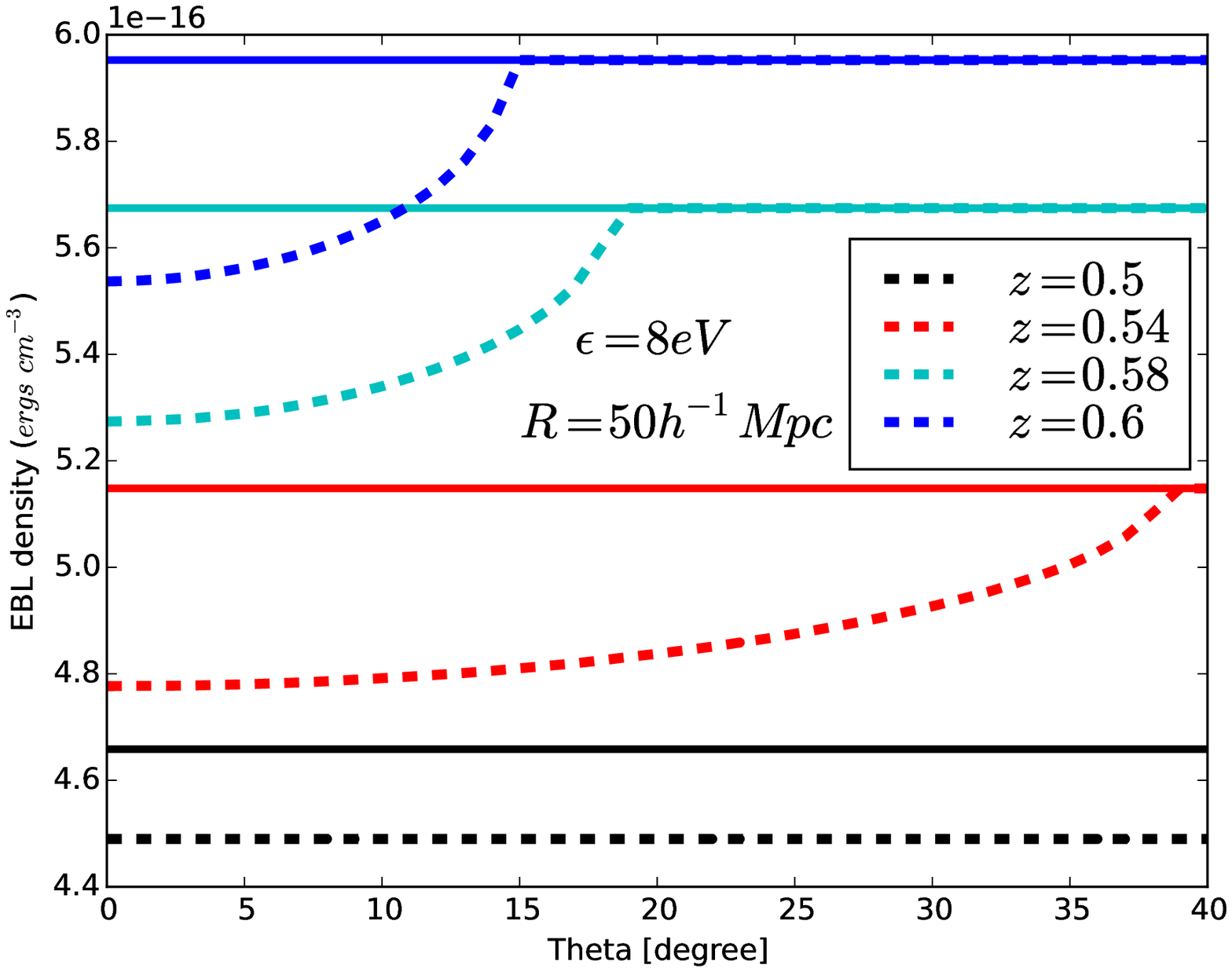}
\caption{Left panel: Relative differential EBL-energy-density deficit as a function of distance along the line
of sight, for various EBL photon energy densities, in the case of a $R = 50 \, h^{-1}$~Mpc void located at
$z_v = 0.5$. The red vertical lines indicate the boundaries of the void. 
Right panel: Angle dependence ($\theta$ is the angle with respect to the direct line of sight through 
the center of the void) 
of the EBL energy density in the presence of a $R = 50 \, h^{-1}$~Mpc 
at $z_v = 0.5$ void (dashed lines), compared to the homogeneous case (solid lines, which does not have
any angle dependence), at a representative near-UV photon energy of $\epsilon = 8$~eV, for two positions 
(redshifts) along the line of sight: at the center of the void (black, lower curves), within the void, but
behind its center (red curves), and behind the void (blue and cyan, upper curves).
\label{fig:EBLdensities3} }
\end{figure}

The relative EBL deficit as a function of distance is plotted for various different photon energies in 
the case of the $R = 50 \, h^{-1}$~Mpc void in the left panel of Figure \ref{fig:EBLdensities3}. 
The right panel of Figure \ref{fig:EBLdensities3} illustrates the angle dependence of the EBL in the presence
of a void, compared to the homogeneous case. Right in the center of the void ($z_v = 0.5$ 
in the example studied here), the EBL is isotropic, due to the assumption of a spherical void, but reduced compared
to the homogeneous EBL case. For positions located outside the void, the reduction is present only for EBL
photon arrival directions intersecting the void, as expected.

\begin{figure}[ht]
\plottwo{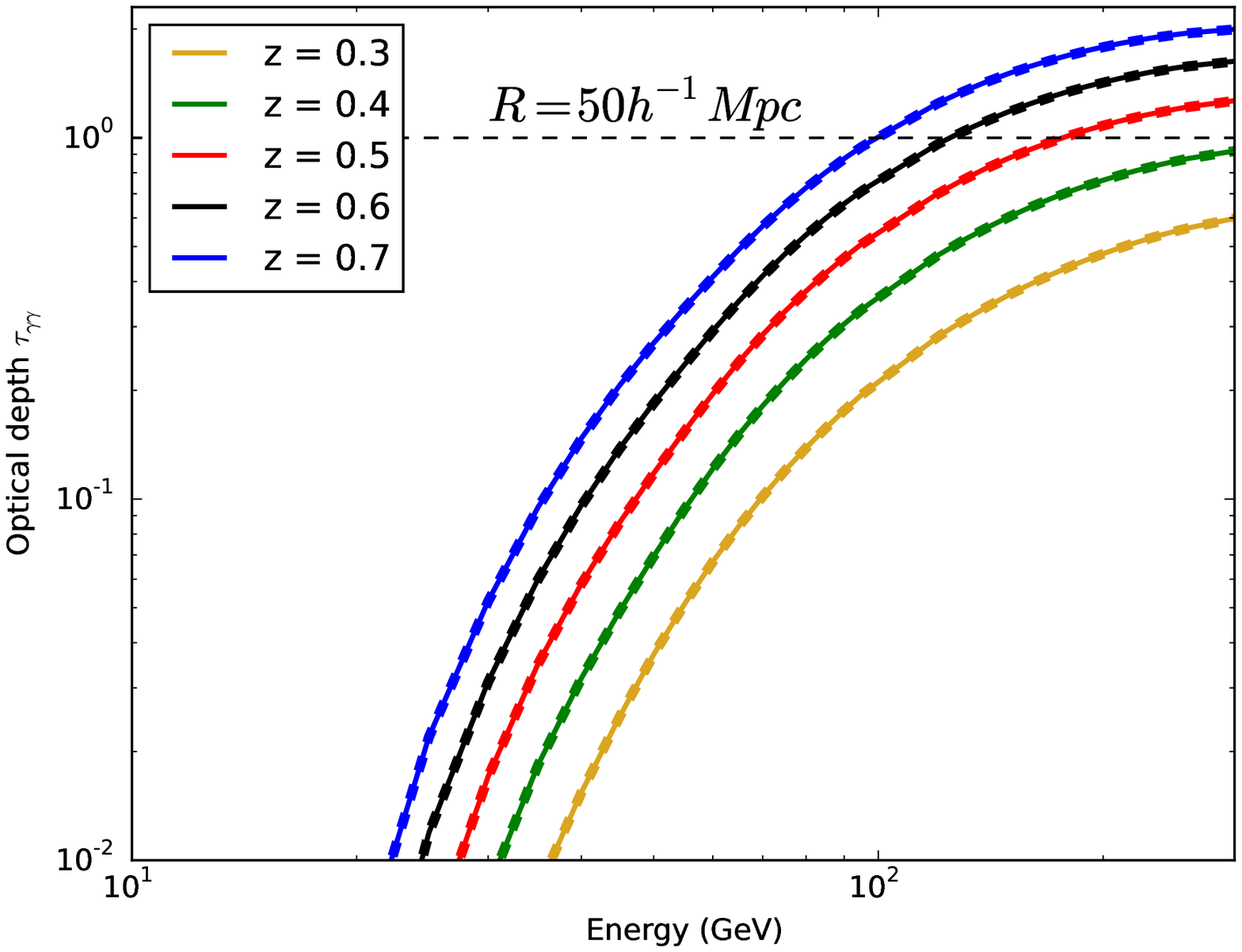}{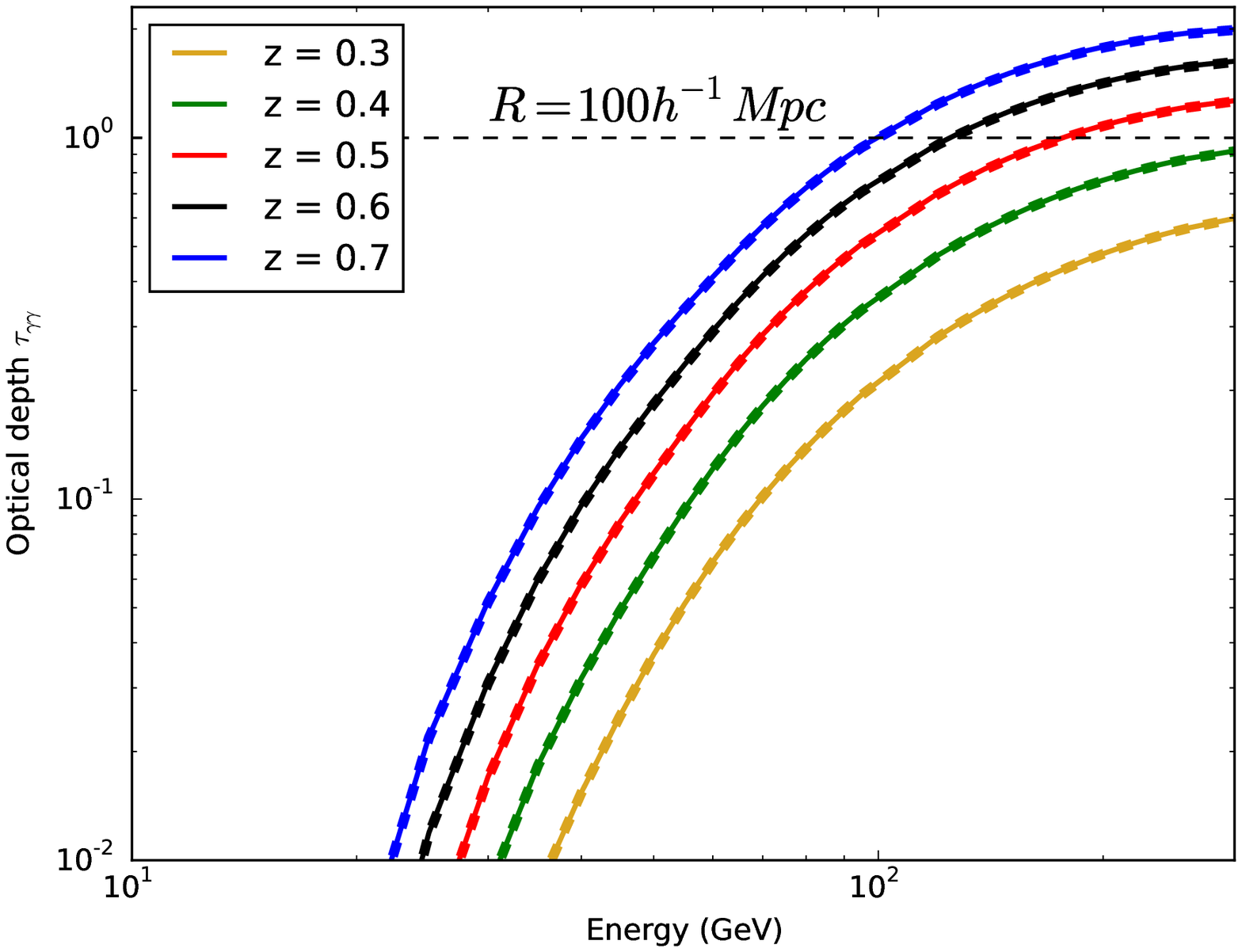}
\plottwo{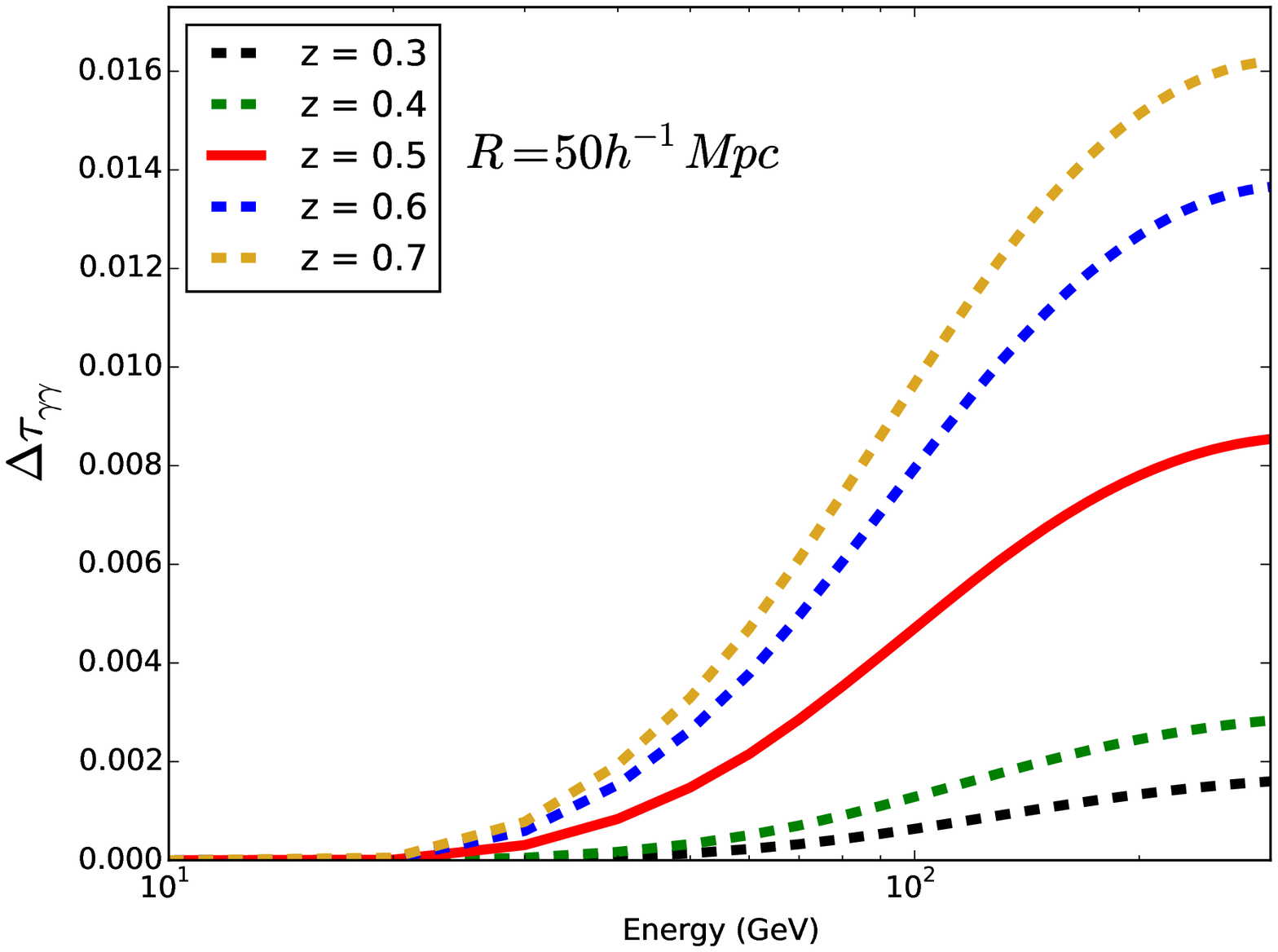}{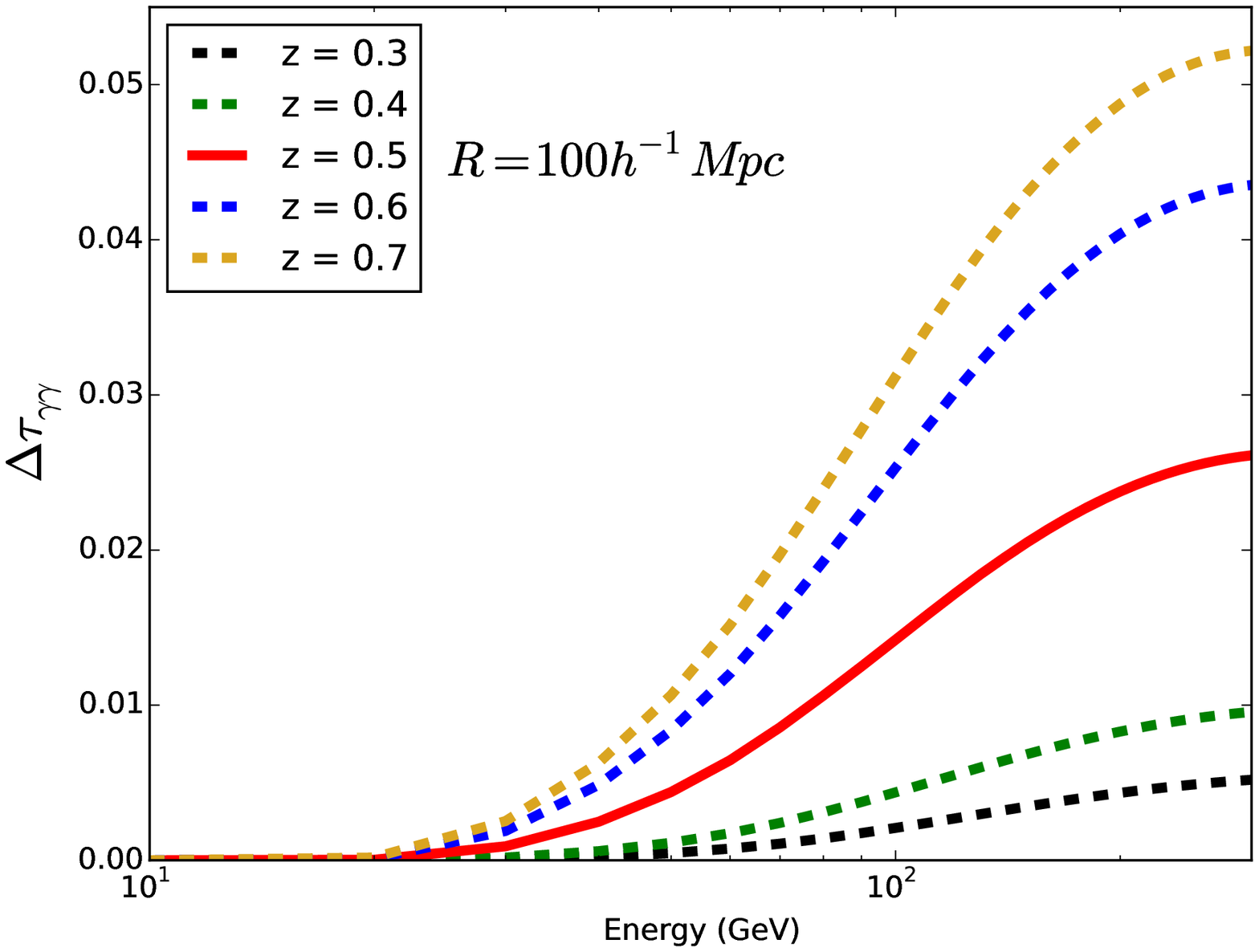}
\caption{Top panels: EBL $\gamma\gamma$ optical depth as a function of $\gamma$-ray photon-energy in the presence 
of a void (dashed), compared to the homogeneous case (solid), for the same example voids as illustrated in
Figure \ref{fig:EBLdensities1}. Bottom panels: $\gamma\gamma$ optical depth deficit due to the presence 
of the voids for the same two cases as in the top panels.
\label{fig:opacity1} }
\end{figure}

Figure \ref{fig:opacity1} illustrates the effect of a void on the $\gamma\gamma$ opacity for the same two 
example cases as illustrated in Figure \ref{fig:EBLdensities1}, for sources located at various redshifts in
front of, within, and behind the void. As expected, the effect is negligible if the source is located in
front of the void (as seen by an observer on Earth), and is maximum for source locations right behind the
void. However, even in the case of the $R = 100 \, h^{-1}$~Mpc void, the maximum effect on the $\gamma\gamma$
opacity is less than 1~\%. Note that the effect on the $\gamma\gamma$ opacity is much smaller than the maximum
EBL energy density deficit in the center of the void due to the integration over the entire line of sight.

\subsection{Multiple voids along the line of sight}
\label{Mvoid}

After investigating the effect of one single cosmic void along the line of sight we now investigate the 
more realistic case of several voids along (or near) the line of sight. From Figure \ref{fig:EBLdensities2}, 
we notice that the relativie EBL-energy-density-deficit scales approximately proportional to the size of 
the void. We therefore conclude that the effect of a number $n$ of voids of radius $R_1$ is aproximately 
the same as the effect of a large void with radius $R_n = n \, R_1$. As a test case, we therefore consider
void sizes up to $R \lesssim 1 \, h^{-1}$~Gpc which approximates the effect of $\lesssim 10$ voids with realistic 
void sizes $R \lesssim 100 \, h^{-1}$~Mpc distributed along or very close to the line of sight. The center of 
the cumulative void is assumed to be located at a redshift of $z_v = 0.3$, considering a source located at 
redshift $z_s \ge 0.6$.\\

\begin{figure}[h]
\plottwo{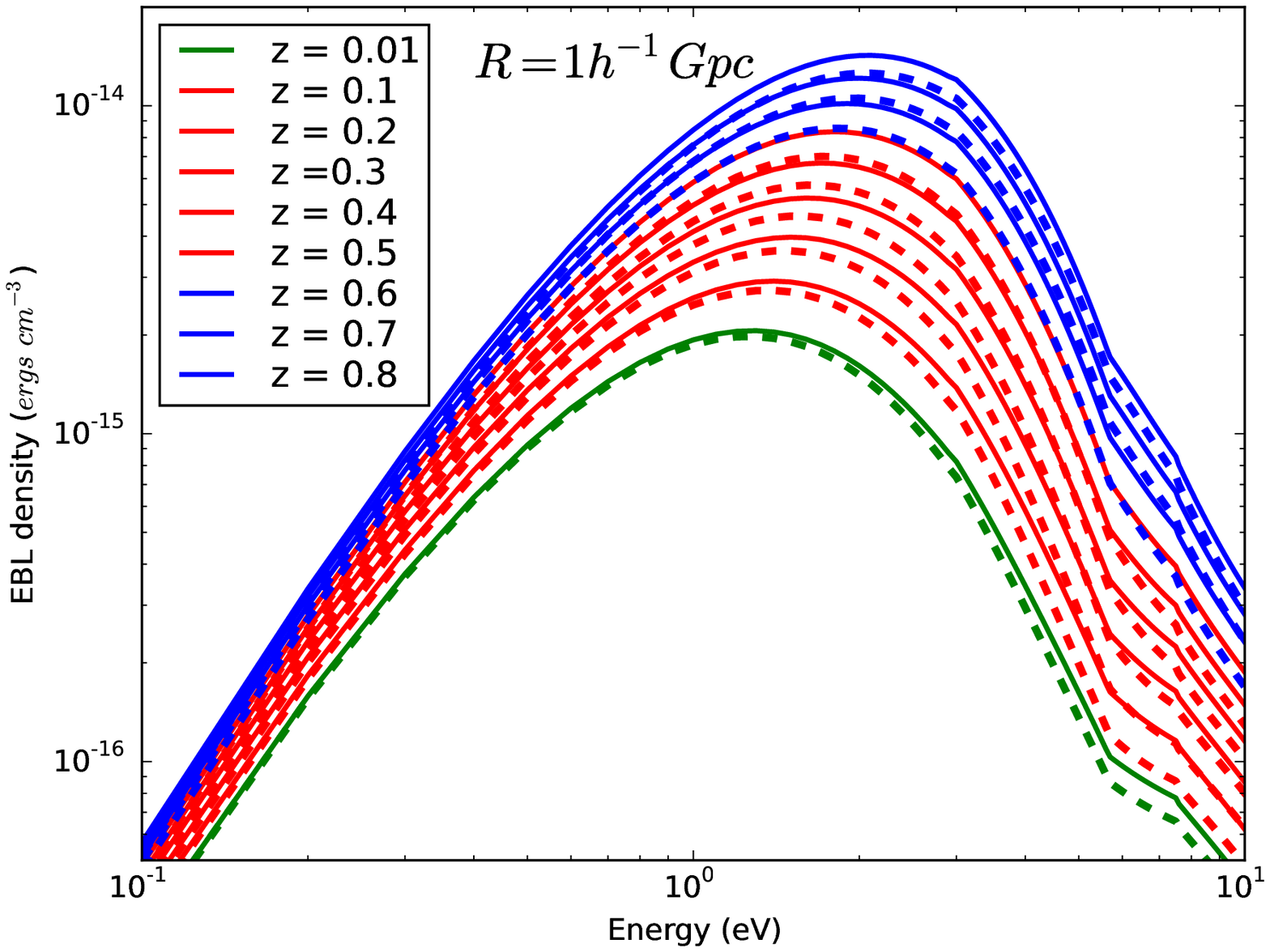} {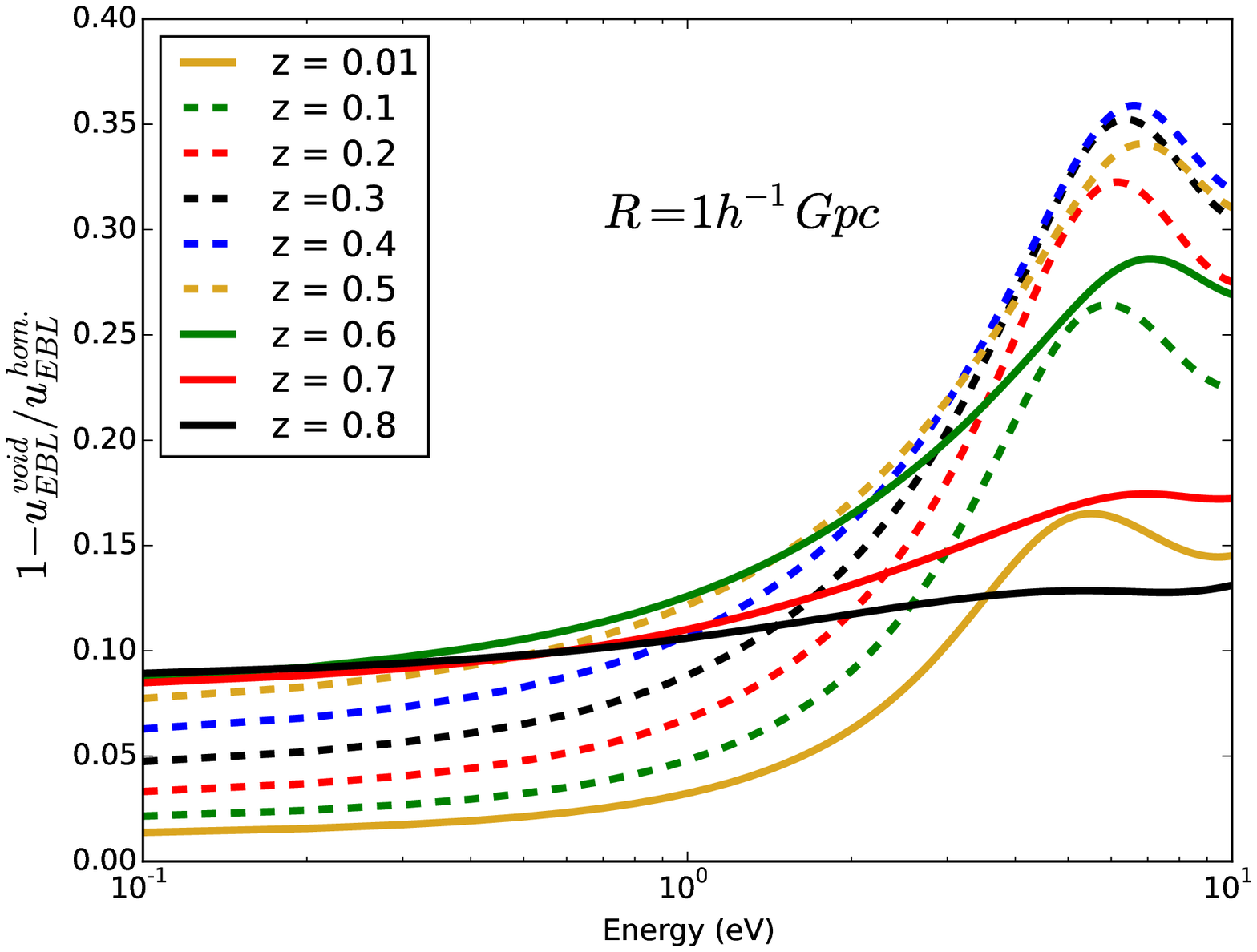}
\plottwo{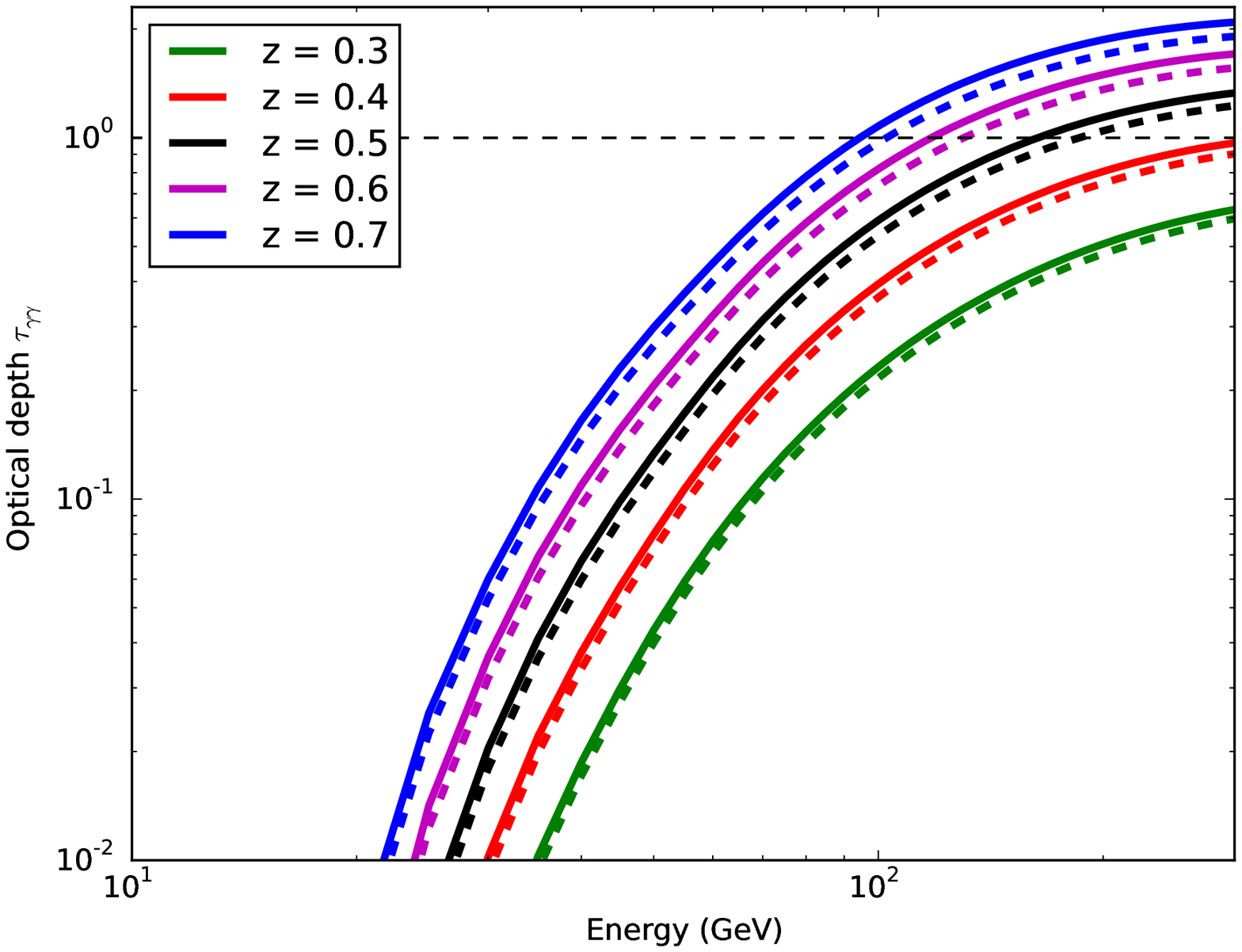}{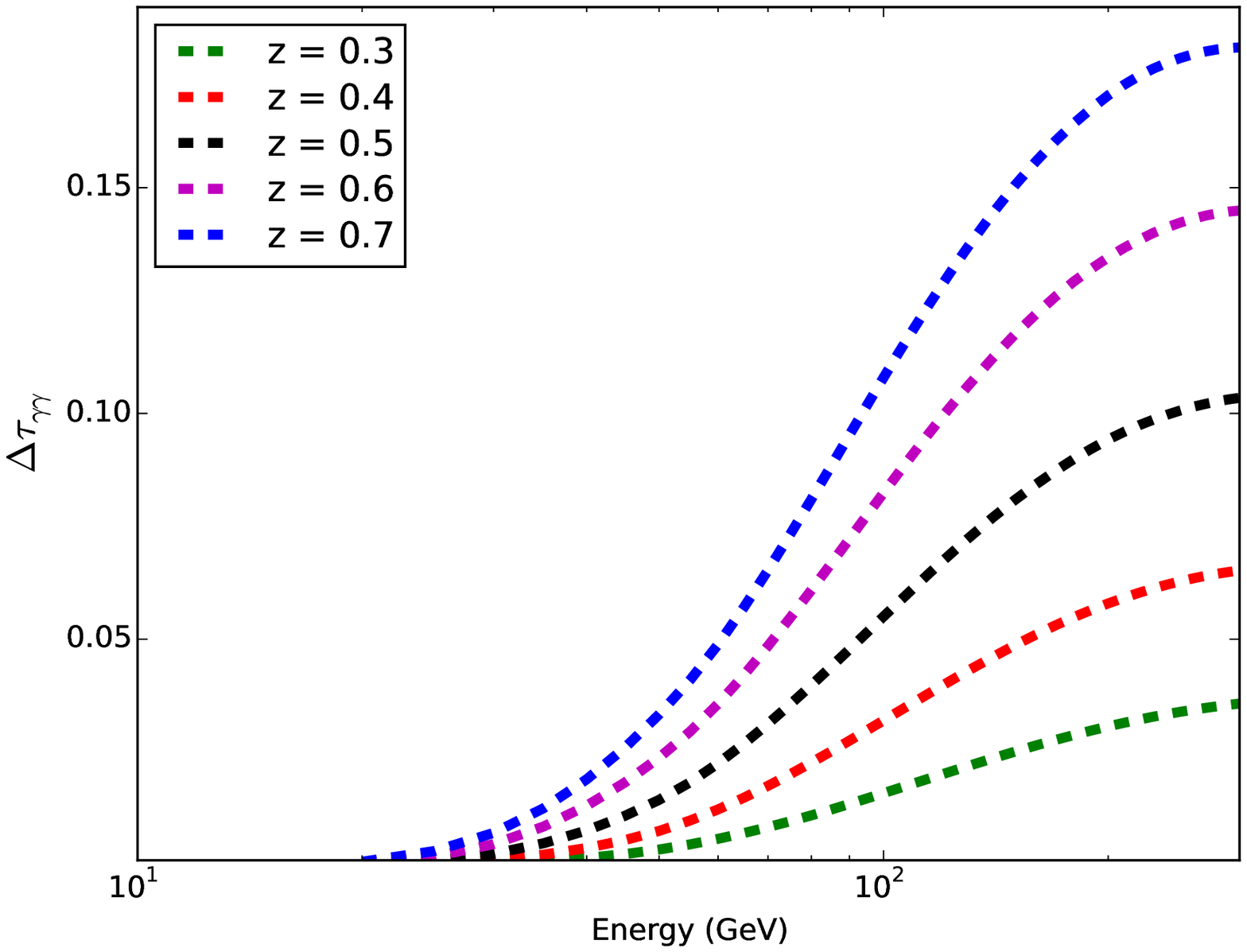}
\caption{Top left panel: Angle-averaged EBL photon energy density spectra for a homogeneous EBL (solid lines) 
and in the presence of an accumulation of 10 cosmic voids of radius $R = 100 \, h^{-1}$~Mpc each (dahed lines),
whose distribution along the line of sight is centered at redshift $z_v = 0.3$. 
Green curves indicate locations in front of the ensemble of voids, red within, and blue behind the ensemble
of voids. Top right panel: Relative deficit of the EBL energy density due to the voids, dash lines and solid lines 
represents the effect of voids to the EBL-energy-density inside and outside the ensemble of voids, respectively.
Bottom left panels: EBL $\gamma\gamma$ {optical depth} as a function of $\gamma$-ray photon-energy in the presence 
of voids (dashed), compared to the homogeneous case (solid), for the same example voids as illustrated in
the top panels. Bottom right panels: $\gamma\gamma$ {optical depth} deficit due to the presence 
of the voids for the same two cases as in the left panel.
\label{fig:EBLsuper} }
\end{figure}

Figure \ref{fig:EBLsuper} (top left panel) compares the EBL energy density spectrum for the case of such 
an accumulation of voids (dashed lines) to the homogeneous EBL case (solid lines). The top right panel of 
Figure \ref{fig:EBLsuper} shows the fractional difference between the homogeneous and the inhomogeneous 
case as a function of photon energy for various redshifts along the line of sight. In the bottom left panel 
of Figure \ref{fig:EBLsuper}, we compare the resulting $\gamma\gamma$ opacities for the case of an ensemble
of voids (dashed lines) and the homogeneous EBL case (solid lines), and the bottom right panel shows the 
$\gamma\gamma$ optical depth deficit due to the presence of the voids for the same two cases as in the 
left panel.
\\
We can notice that for the extreme case of an accumulation of about 10 voids of typical sizes along the line 
of sight to a blazar, the EBL energy density even at the center of the cumulative void is reduced by around 35~\%, 
and the resulting maximum $\gamma\gamma$ opacity reduced by around 15~\%. This is because even if the star-formation 
rate is set to zero within the void, the EBL density within the void is still substantial due to the contributions 
from the rest of the Universe outside the void.

\subsection{Application to PKS 1424+240}
\label{PKS1424}

\cite{Furniss15} had investigated the possibility that the unusually hard VHE $\gamma$-ray spectra observed
in some distant ($z \gtrsim 0.5$) $\gamma$-ray blazars might be due to a reduced EBL density caused by
galaxy underdensities along the line of sight. Specifically, they investigated a possible correlation between
galaxy-count underdensities based on the Sloan Digital Sky Survey (SDSS) and the positions of hard-spectrum
VHE blazars, and found a tentative hint for such a correlation (although the small sample size prevented the
authors from drawing firm conclusions). Based on this result, as a first estimate of the effect of such 
underdensities on the EBL, they suggested a linear scaling of the line-of-sight galaxy density with the 
EBL $\gamma\gamma$ opacity. For the specific case of the distant \citep[$z \ge 0.6$,][]{Furniss13} VHE blazar 
PKS 1424+240, they found that the reduction of the EBL resulting from such a direct linear scaling is not 
sufficient to remove the apparent spectral hardening of the VHE $\gamma$-ray spectrum observed by VERITAS, 
when correcting for EBL absorption based on state-of-the-art (homogeneous) EBL models \citep{Archambault14}.

\begin{figure}[ht]
\plottwo{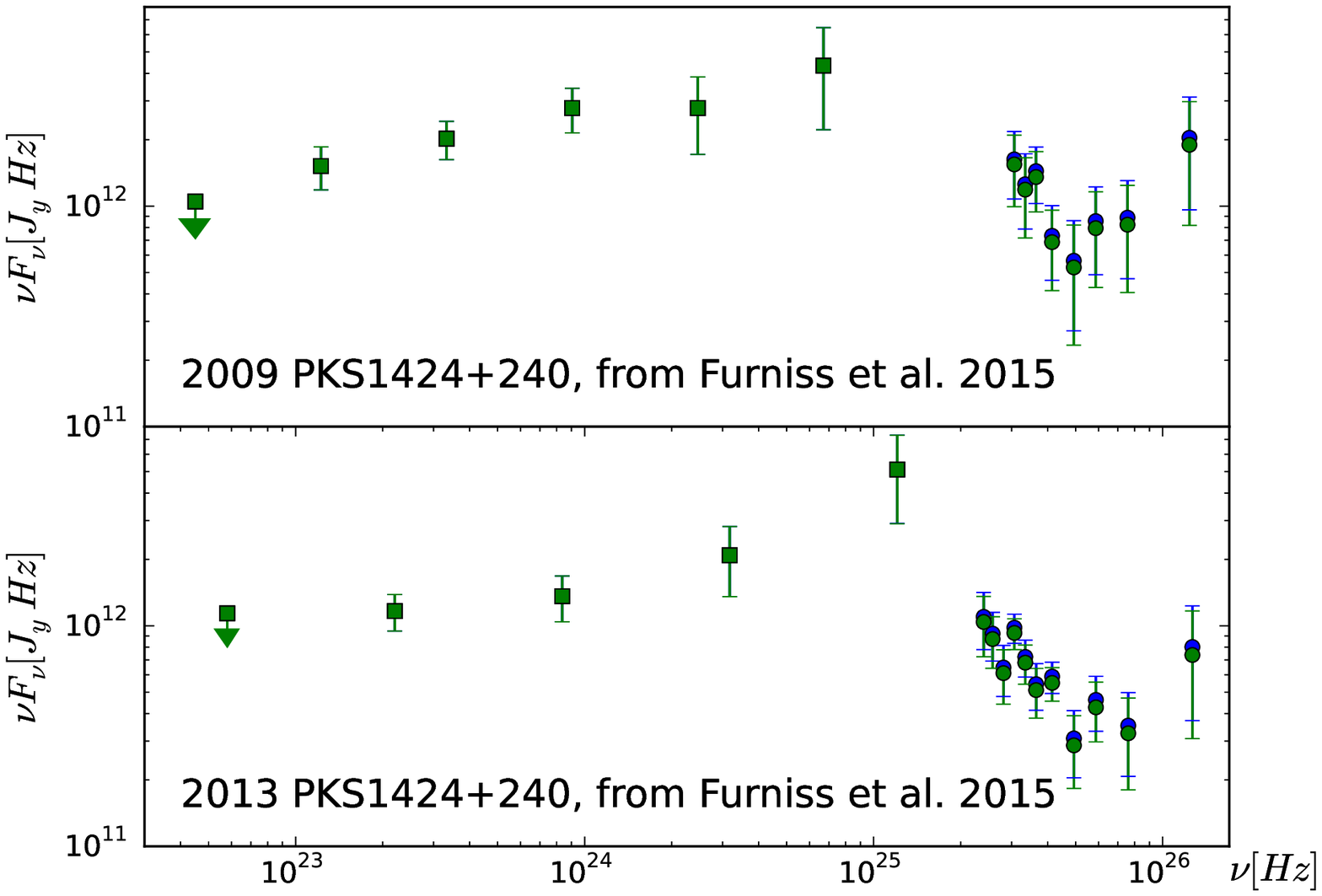}{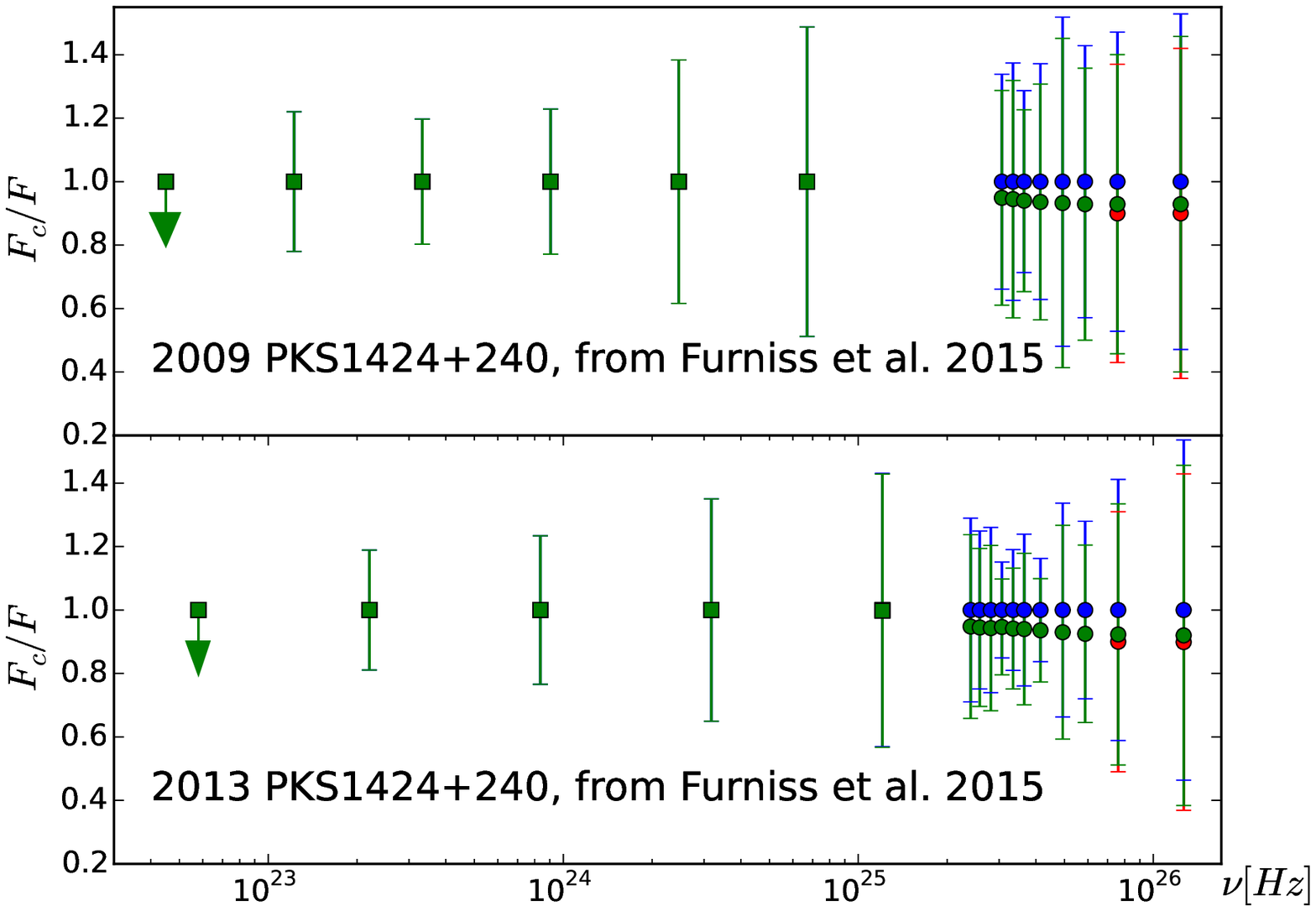}
\caption{HE -- VHE $\gamma$-ray spectra of PKS1424+240, from \cite{Archambault14}. The blue points
show the EBL-corrected spectrum using the \cite{Gilmore12} EBL model. The red points represent the reduced
EBL correction, using the linear scaling of $\tau_{\gamma\gamma}$ with the line-of-sight galaxy density,
as suggested by \cite{Furniss15}. The green points illustrate the reduced EBL correction resulting from
our model calculation with a void of radius $R = 50h^{-1}$ Mpc centered at $z_v = 0.5$ (comoving distance 
$1.724$ Gpc), assuming a source redshift of $z_s = 0.6$ (comoving distance $2.056$ Gpc), which results in
approximately the same perceived line-of-sight galaxy underdensity as used by \cite{Furniss15}. The left
panels show the actual $\nu F_{\nu}$ spectra, while the right panels show the spectra normalized to the
EBL-corrected flux points from \cite{Archambault14}.
\label{fig:Furniss} }
\end{figure}

For an assumed redshift of $z = 0.6$ for PKS~1424+240, our example case of $R = 50 \, h^{-1}$~Mpc and $z_v = 0.5$ 
results in approximately the same galaxy-count underdensity factor as found by \cite{Furniss15} along the line of
sight to this source. In Figure \ref{fig:Furniss}, we therefore compare the EBL reduction effect based on
the direct linear scaling with galaxy underdensity, with our detailed EBL calculation assuming a void
along the line of sight, for the two observing periods presented in \cite{Archambault14}. The left panel
illustrates the effect on the actual $\nu F_{\nu}$ spectra, while the right panel shows the $\gamma$-ray
spectra normalized to the flux points corrected by the homogeneous \cite{Gilmore12} EBL-absorption model. 
The figure illustrates that the EBL-opacity reduction effect due to the void, found in our detailed
calculations, is slightly smaller than the effect resulting from a direct linear scaling with galaxy
underdensity. Thus, we conclude that the tentative spectral hardening of the VHE spectrum of PKS~1424+240 
is likely not an artifact of an under-estimate of the EBL opacity due to possible EBL inhomogeneities.

\section{Summary and Conclusions}
\label{Summary}

We have presented detailed calculations of the effect of cosmic inhomogeneities on the EBL and the
resulting $\gamma\gamma$ opacity for VHE $\gamma$-ray photons from sources at cosmological distances.
Specifically, we have considered the presence of a cosmic void, which, for simplicity, we have 
represented as a spherical region in which the local star formation rate is zero. We have shown 
that for realistic void sizes of $R \lesssim 100 \, h^{-1}$~Mpc, the EBL energy density even at
the center of the void is reduced by less than 10~\%. Even if the void is located right in front
of the background $\gamma$-ray source, the $\gamma\gamma$ opacity is reduced by typically less 
than 1~\%.  We found an approximately linear scaling of the EBL deficit effect with the size of the void.
Even in the presence of a large number of voids adding up to a total line of sight distance through 
voids of $\sim 2 \, h^{-1}$~Gpc ($= 2 \, R_n$), the EBL $\gamma\gamma$ opacity is only reduced 
by $\sim 15$~\%.

This reduction is smaller than obtained from a direct linear scaling of the $\gamma\gamma$
opacity with galaxy-count underdensities along the line of sight to $\gamma$-ray sources. For the 
specific case of PKS 1424+240, we have illustrated that the inferred (marginal) spectral hardening 
of the VHE $\gamma$-ray spectrum, after correction for EBL absorption, 
if confirmed by future, more sensitive VHE $\gamma$-ray observations,
is most likely not an artifact of an over-estimation of the EBL opacity due to cosmic inhomogeneities. 

{Since we have shown that realistic EBL inhomogeneities do not lead to a significant reduction 
of the EBL $\gamma\gamma$ opacity, hints for unexpected spectral hardening of the VHE spectra of 
several blazars remain, for which other explanations would have to be invoked, if they can be 
confirmed by future observations (e.g., by the Cherenkov Telescope Array, CTA).
One possibility is that this hardening
is, in fact, a real, intrinsic feature of the $\gamma$-ray spectra of these blazars, possibly
due to a pion-production induced cascade component in a hadronic blazar model scenario 
\citep[e.g.,][]{Boettcher13,Cerruti15}. If such a spectral hardening is not intrinsic
to the source, more exotic explanations, such as ALPs or a cosmic-ray induced secondary
radiation component, would need to be invoked.

\section{Acknowledgments}

We thank Amy Furniss for stimulating discussions and for sharing the PKS 1424+240 data with us. 
We also thank the anonymous referee for a careful reeding of the manuscript and helpful 
suggestions which have led to significant improvements of the manuscript.
The work of M.B. is supported through the South African Research Chair Initiative of the National 
Research Foundation\footnote{Any opinion, finding and conclusion or recommendation expressed in 
this material is that of the authors and the NRF does not accept any liability in this regard.} 
and the Department of Science and Technology of South Africa, under SARChI Chair grant No. 64789.



\end{document}